\documentclass[%
 reprint,
superscriptaddress,
 amsmath,amssymb,
 aps,
prb,
]{revtex4-2}

\usepackage{color}
\usepackage{graphicx}
\usepackage{dcolumn}
\usepackage{bm}
\usepackage{hyperref}
\hypersetup{colorlinks=true,linkcolor=blue,citecolor=blue,urlcolor=blue}

\usepackage[T1]{fontenc} 
\usepackage{newtxtext}   
\usepackage{newtxmath}   
\usepackage{upgreek}

\begin{document}

\preprint{APS/123-QED}

\title{Anomalous Nernst effect in a ferrimagnetic nodal-line semiconductor Mn$_3$Si$_2$Te$_6$}  

\author{Chen Ran}
\thanks{These authors contributed equally to this work. }
\affiliation{College of Physics \& Center of Quantum Materials and Devices, Chongqing University, Chongqing 401331, China}

\author{Xinrun Mi}
\thanks{These authors contributed equally to this work. }
\affiliation{College of Physics \& Center of Quantum Materials and Devices, Chongqing University, Chongqing 401331, China}

\author{Junying Shen}
\thanks{These authors contributed equally to this work. }
\affiliation{Institute of High Energy Physics, Chinese Academy of Sciences (CAS), Beijing 100049, China}
\affiliation{Spallation Neutron Source Science Center, Dongguan 523803, China}

\author{Honghui Wang}
\affiliation{College of Physics \& Center of Quantum Materials and Devices, Chongqing University, Chongqing 401331, China}

\author{Kunya Yang}
\affiliation{College of Physics \& Center of Quantum Materials and Devices, Chongqing University, Chongqing 401331, China}

\author{Yan Liu}
\affiliation{Analytical and Testing Center, Chongqing University, Chongqing 401331, China}

\author{Guiwen Wang}
\affiliation{Analytical and Testing Center, Chongqing University, Chongqing 401331, China}

\author{Guoyu Wang}
\affiliation{College of Physics \& Center of Quantum Materials and Devices, Chongqing University, Chongqing 401331, China}

\author{Youguo Shi}
\affiliation{Beijing National Laboratory for Condensed Matter Physics and Institute of Physics, Chinese Academy of Sciences, Beijing 100190, China}

\author{Aifeng Wang}
\affiliation{College of Physics \& Center of Quantum Materials and Devices, Chongqing University, Chongqing 401331, China}

\author{Yisheng Chai}
\affiliation{College of Physics \& Center of Quantum Materials and Devices, Chongqing University, Chongqing 401331, China}

\author{Xiaolong Yang}
\affiliation{College of Physics \& Center of Quantum Materials and Devices, Chongqing University, Chongqing 401331, China}

 \author{Mingquan He}
\email{mingquan.he@cqu.edu.cn}
\affiliation{College of Physics \& Center of Quantum Materials and Devices, Chongqing University, Chongqing 401331, China}

\author{Xin Tong}
\email{tongx@ihep.ac.cn}
\affiliation{Institute of High Energy Physics, Chinese Academy of Sciences (CAS), Beijing 100049, China}
\affiliation{Spallation Neutron Source Science Center, Dongguan 523803, China}

\author{Xiaoyuan Zhou}
\email{xiaoyuan2013@cqu.edu.cn}
\affiliation{College of Physics \& Center of Quantum Materials and Devices, Chongqing University, Chongqing 401331, China}

\date{\today}

\begin{abstract}
In the ferrimagnetic nodal-line semiconductor Mn$_3$Si$_2$Te$_6$, colossal magnetoresistance (CMR) arises below $T_\mathrm{c}=78$ K due to the interplay of magnetism and topological nodal-line fermiology. The Berry curvature associated with the topological nodal-line is expected to produce an anomalous Nernst effect. Here, we present sizable anomalous Nernst signal in Mn$_3$Si$_2$Te$_6$ below $T_\mathrm{c}$. In the low-magnetic-field region where CMR is most apparent, the scaling ratio between the Nernst signal and magnetization is significantly enhanced compared to that in conventional magnetic materials. The enhanced Nernst effect and CMR likely share the same mechanisms, which are closely linked to the nodal-line topology.

\end{abstract}

\maketitle

In magnetic topological materials, the interplay of magnetism and band topology often gives rise to a variety of nontrivial transport properties, such as 
giant anomalous Hall effect (AHE), giant anomalous Nernst effect (ANE) and quantum anomalous Hall effect \cite{Tokura2019,ikhlas2017large,Liu2018,ZhangyuanboMnBiTe, Yin2020,Nakatsuji2022}. The intimate coupling between spin degrees of freedom and electronic band topology allows easy manipulation of spin or charge transport by tailoring spin configurations using external magnetic fields. The peculiar transport effects together with their easy tunability in magnetic topological materials are highly favored in developing magnetoelectric, spintronic device applications.    

\begin{figure*}
\includegraphics[width=450pt]{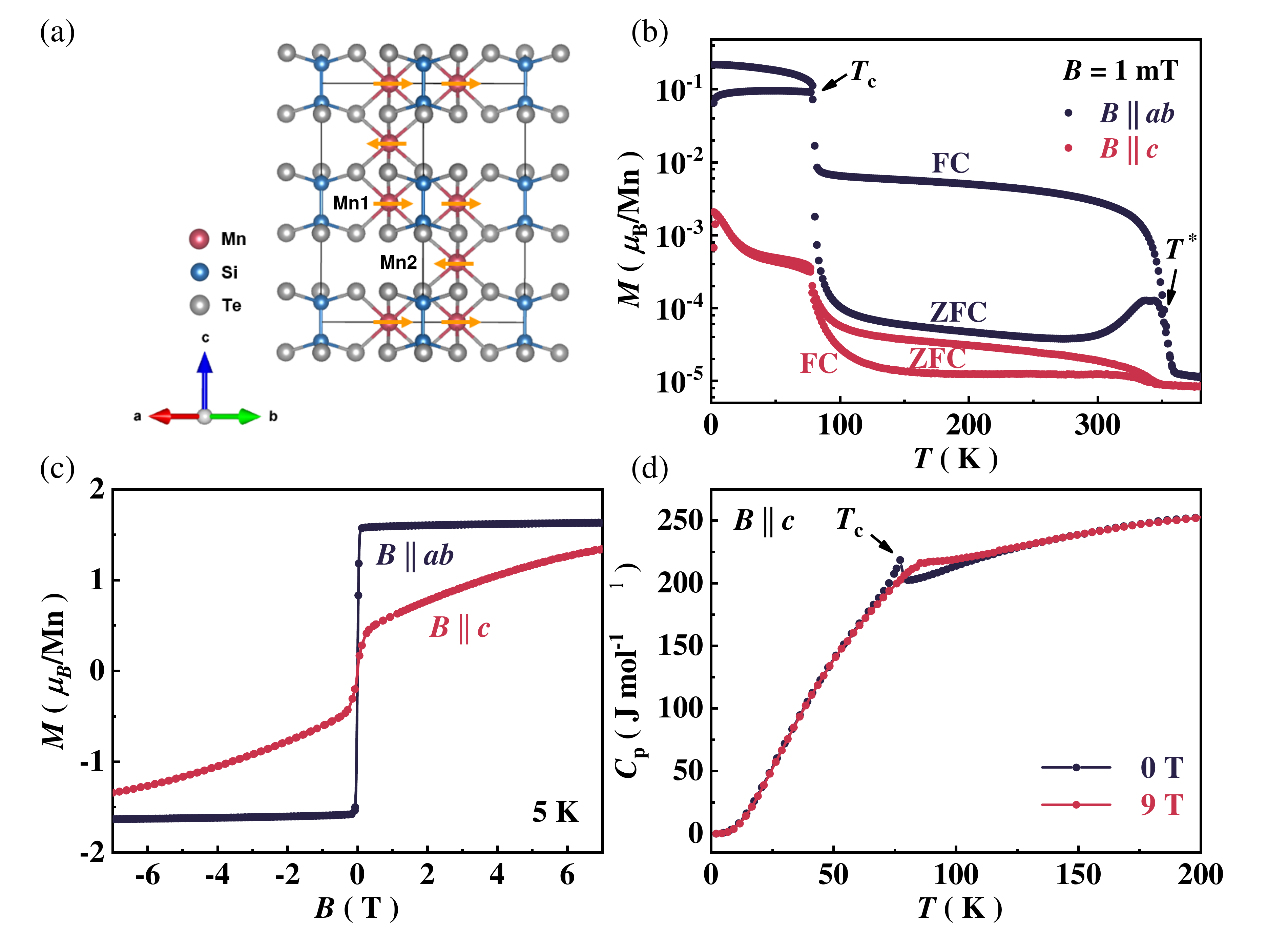}
\caption{\label{fig:1}(a) Crystal structure of Mn$_3$Si$_2$Te$_6$ (side view). Arrows on Mn atoms represent magnetic moments which order ferrimagnetically below $T_\mathrm{c}=78$ K. (b) Temperature dependence of magnetization $M(T)$ for a Mn$_3$Si$_2$Te$_6$ single crystal. Both zero-field-cooled (ZFC) and field-cooled (FC) results with a magnetic field $B$ = 1 mT applied within the $ab$-plane and along the $c$-axis are shown. In addition to the ferrimagnetic transition at $T_\mathrm{c}$, another transition is also seen at $T^*$. (c)  Isothermal magnetization $M(H)$ measured at $T$ = 5 K. (d) Temperature-dependent specific heat $C_\mathrm{p}(T)$ captured in 0 and 9 T.} 
\end{figure*}

Of particular interest is the ferrimagnetic nodal-line semiconductor Mn$_3$Si$_2$Te$_6$, which displays colossal magnetoresistance (CMR) in the ferrimagnetic state \cite{ni2021colossal,zhang2022control,seo2021colossal,wang2022pressure}. As shown in Fig.~\ref{fig:1}(a), Mn$_3$Si$_2$Te$_6$ is a van der Waals layered material, which crystallizes in a trigonal structure (space group: $P$$\bar{3}$1$c$, No. 163). There are two types of Mn atoms. The Mn1 atoms and Te atoms form edge sharing 
MnTe$_6$ octahedra, creating honeycomb layers within the $ab$ plane. The Mn2 atoms are sandwiched by the Mn1 layers, forming a triangular pattern. Below $T_\mathrm{c}\sim$ 78 K, Mn$_3$Si$_2$Te$_6$ enters a ferrimagnetic state, in which ferromagnetic Mn1 layers and Mn2 layers are bonded antiferromagnetically [see Fig.~\ref{fig:1}(a)] \cite{may2017magnetic}. Magnetic moments lie within the $ab$ plane with an easy-plane anisotropy. In the ferrimagnetic state, negative CMR reaching up to nine orders of magnitude appears only when an external magnetic field is directed along the magnetic hard $c$-axis, in contrast to conventional CMR materials \cite{ni2021colossal,zhang2022control,seo2021colossal}. Only moderate negative magnetoresistance is found when a magnetic field is applied within the magnetic easy $ab$-plane, leading to colossal angular magnetoresistance up to $\sim10^{11}$\% per radian \cite{seo2021colossal}. It has been suggested that the CMR arises from a metal-insulator transition, which is induced by lifting the spin orientation-dependent nodal-line band degeneracy \cite{seo2021colossal}. Alternatively, chiral orbital currents (COC) may emerge within the $ab$-plane, producing orbital magnetic moments along the $c$-axis \cite{zhang2022control}. These orbital magnetic moments and COC domains interact strongly with the Mn moments, giving rise to negative CMR by aligning the COC domains in a $c$-axis magnetic field \cite{zhang2022control}. Still, the nature of the observed unusual CMR in Mn$_3$Si$_2$Te$_6$ remains elusive.  

When the topological nodal-line degeneracy is lifted, finite Berry curvature arises near the Fermi level \cite{seo2021colossal}. Consequently, other intriguing transport properties may also appear, such as the AHE and ANE \cite{xiao2010berry,xiao2006berry}. In particular, the ANE is extremely sensitive to the Berry curvatures close to the Fermi energy ($E_F$) as the ANE probes the energy derivative of charge transport at $E_F$ \cite{xiao2010berry,xiao2006berry}. Indeed, Berry curvature-induced large ANE  has been observed in various magnetic topological semimetals, such as  Co$_3$Sn$_2$S$_2$ \cite{guin2019zero,CoSnS,DingCoSnS}, Co$_2$MnGa \cite{guin2019anomalous,XuCoMnGa,Sakai2018CoMnGa}, Mn$_3$Sn \cite{ikhlas2017large,XiaokangMn3Sn}, Fe$_3$Sn$_2$ \cite{zhang2021Topological} and Fe$_3$GeTe$_2$ \cite{xu2019large}. In this report, we probe the Berry curvature associated with the nodal-line fermiology in Mn$_3$Si$_2$Te$_6$ using thermoelectric Seebeck and Nernst measurements. Large magneto-Seebeck effect and ANE are found below $T_\mathrm{c}$ in magnetic fields applied along the $c$-axis. Importantly, in small magnetic fields, the ratio between the Nernst signal and magnetization strongly deviates from that in conventional magnets, pointing to the important roles played by the Berry curvature associated with the nontrivial band topology.

Single crystals of Mn$_3$Si$_2$Te$_6$ were grown using the flux method \cite{seo2021colossal}. Starting materials of Mn, Si and Te were mixed in a molar ratio of 3:2:12 in Ar atmosphere. The mixed materials were then placed in an alumina crucible and sealed in an evacuated quartz tube. The tube was first heated in a muffle furnace to 1000 $^{\circ}$C and dwelt for 20 h. Then the furnace was slowly cooled to 750 $^{\circ}$C at a rate of 1.5 $^{\circ}$C/h. The tube was quickly taken out of the furnace, and Mn$_3$Si$_2$Te$_6$ single crystals with typical dimensions of 5 $\times$ 5 $\times$ 1 mm$^3$ were finally obtained after removing the flux by centrifugation.

The crystal structure of Mn$_3$Si$_2$Te$_6$ crystals was characterized by single crystal diffraction experiments using a Bruker D8 Venture diffractometer (see Supplemental Material \cite{supplemental}).  The magnetic properties were measured in a magnetic properties measurement system (MPMS-3, Quantum Design Inc.). Resistivity, Hall resistivity were performed in a physical property measurement system (PPMS Dyna cool 9 T, Quantum Design Inc.) using a standard Hall-bar geometry. The longitudinal and transverse thermoelectric measurements were carried out in a cryostat (TESPT14T50, Oxford Instruments inc.) using a home-built setup equipped with one heater and two thermometers.

\begin{figure*}
\includegraphics[width=460pt]{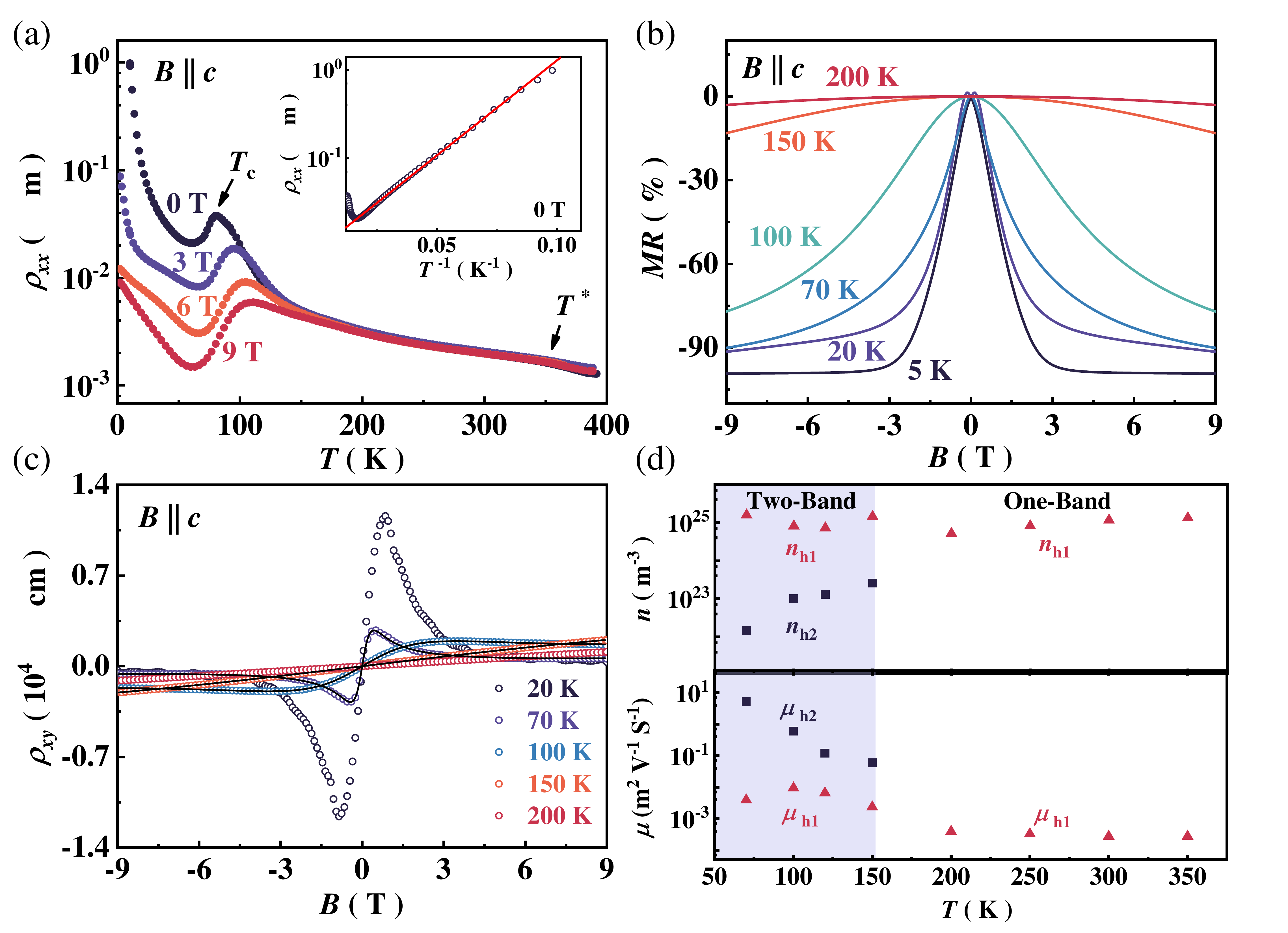}
\caption{\label{fig:2}(a) Temperature dependence of in-plane resistivity $\rho_{xx}$ measured with magnetic fields applied along the $c$-axis. The inset in (a) shows the fitting (red line) of resistivity between 150 K and 300 K using a thermally activated model. (b) The magnetoresistance $\mathit{MR} = [\rho _{xx}(\mathrm{B} )-\rho _{xx}(0)]/\rho _{xx}(0)$ measured at different temperatures with $\mathit{B} \parallel \mathit{c}$. (c) The Hall resistivity $\rho _{xy}$ recorded at selective temperatures. Solid black lines are theoretical description using a two-band model. (d) Estimated carrier density ($n$) and mobility ($\mu$) obtained from the analysis of Hall resistivity. }
\end{figure*}

 In Fig.~\ref{fig:1}(b), we present the temperature dependence of magnetization for a Mn$_3$Si$_2$Te$_6$ single crystal. The ferrimagnetic transition is seen clearly at $T_\mathrm{c}=78$ K, agreeing well with other reports \cite{may2017magnetic,ni2021colossal,seo2021colossal,zhang2022control,wang2022pressure}. Below $T_\mathrm{c}$, the in-plane magnetization ($B$ $\|$ $ab$) is 2 orders of magnitude larger than that of out-of-plane ($B$ $\|$ $c$), in accordance with an in-plane easy axis [see Fig. \ref{fig:1}(a)]. Note that this easy-plane anisotropy appears already above $T_\mathrm{c}$ and persists up to $T^*\sim$ 350 K above which the paramagnetic state is recovered. The transition at $T^*$ is consistently seen in melt-grown crystals \cite{may2017magnetic,liu2021polaronic,ni2021colossal}, but disappears in samples prepared by the vapor transport method \cite{may2020magnetic}. The nature of the transition at $T^*$ is still not clear, and may originate from short-range ordering \cite{may2017magnetic}. The strong magnetic anisotropy in the ferrimagnetic state can be further revealed by isothermal magnetization measurements, as shown in Fig.~\ref{fig:1}(c).  The magnetization saturates rapidly to $\sim$ 1.6 $\mu$$_\mathrm{B}$/Mn in a weak magnetic field of  0.1 T for $B$ $\|$ $ab$. For $B$ $\|$ $c$, on the other hand, a strong magnetic field up to  7 T can not fully polarize the magnetic moments (see more data in Supplemental Material \cite{supplemental}). The field anisotropy is found to be as large as 13 T \cite{ni2021colossal}.

The temperature dependence of specific heat ($C_\mathrm{p}$) of Mn$_3$Si$_2$Te$_6$ is plotted in Fig.~\ref{fig:1}(d). In zero magnetic field, the ferrimagnetic transition is manifested as a $\lambda$-shaped peak at $T_\mathrm{c}$.  The transition is smeared out in the presence of magnetic fields and shifts gradually towards higher temperatures due to the suppression of magnetic fluctuations. Above 150 K, $C_\mathrm{p}$ recorded in 0 and 9 T coincides. This indicates that magnetic fluctuations appear well above $T_\mathrm{c}$ and survive to about 150 K. Short-range spin fluctuations may already develop below $T^*$,  as suggested by diffuse neutron scattering experiments \cite{may2017magnetic}.    

\begin{figure*}
\includegraphics[width=450pt]{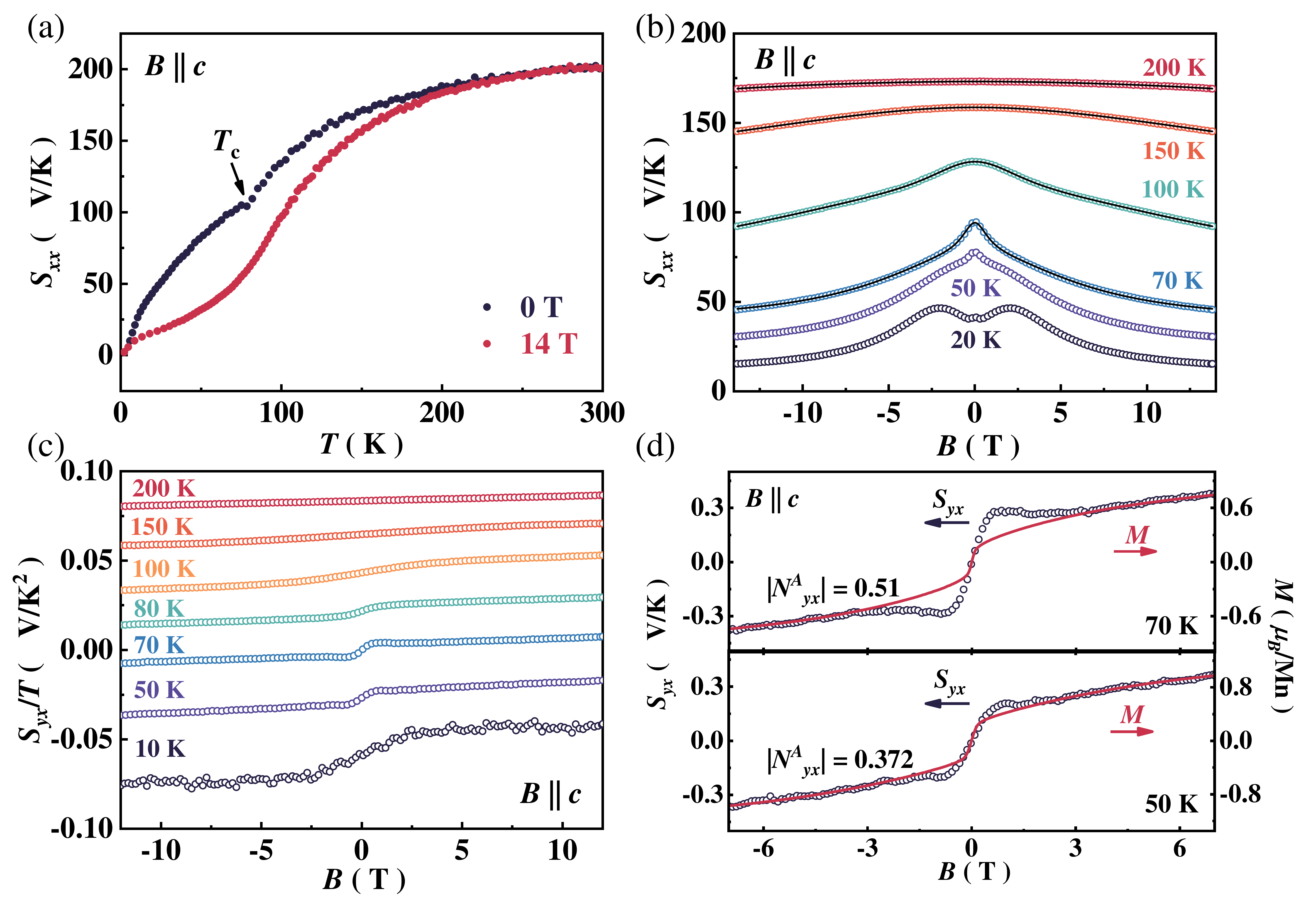}
\caption{\label{fig:3}(a) Temperature dependence of the in-plane Seebeck coefficient $S_{xx}$ measured in $B$ = 0 T and 14 T applied along the $c$-axis. (b) Magneto-Seebeck effect $S_{xx}(B)$ measured at various temperatures. Solid black lines are two-band fitting of the Seebeck signal. (c) The Nernst effect $S_{yx}/T$ measured at fixed temperatures. Curves have been shifted vertically for clarity.  (d) Scaling between the Nernst signal $S_{yx}(B)$ and magnetization $M$. In the low-field region below 3 T, the Nernst effect is enhanced compared to conventional contributions from magnetization. }
\end{figure*}

Figure \ref{fig:2} shows electrical transport results of a Mn$_3$Si$_2$Te$_6$ single crystal.  As shown in Fig.~\ref{fig:2}(a), the in-plane resistivity $\rho_{xx}$ increases rapidly with cooling, pointing to a semiconducting nature. Within 150 - 300 K, $\rho_{xx}(T)$ measured in 0 T follows a thermally activated model $\rho_{xx}(T) = \rho_0\mathrm{exp}(\Delta/k_\mathrm{B}T)$, with an activation energy $\Delta = 24.3(1)$ meV [see the inset in Fig.\ref{fig:2}(a)], in agreement with other reports \cite{seo2021colossal,may2017magnetic}. Here, $k_\mathrm{B}$ is the Boltzmann constant, and $\rho_0$ is a material-dependent constant. Further cooling below 150 K, $\rho_{xx}(T)$ deviates from the simple activation behavior due to increasing scattering of carriers by sizable spin fluctuations. Establishment of long-range ferrimagnetic order below $T_\mathrm{c}$ strongly suppresses magnetic fluctuations, leading to a clear drop in $\rho_{xx}$ when passing through $T_\mathrm{c}$ upon cooling. Note that the transition at $T^*$ appears as a broad hump in resistivity.

The most notable feature of Mn$_3$Si$_2$Te$_6$ is the negative CMR. As seen in Figs.~\ref{fig:2}(a, b), $\rho_{xx}$ is significantly reduced in the presence of out-of-plane magnetic fields below 150 K. The negative CMR becomes particularly evident at lower temperatures. At 5 K,  MR = $[\rho_{xx}(B)-\rho_{xx}$(0 T)]$/\rho_{xx}$(0 T) reaches -99.2\% in 9 T. Note that the CMR is most profound in the low-field region below 3 T. Intriguingly, the CMR only appears when a magnetic field is applied along the magnetic hard axis ($B\parallel c$) \cite{ni2021colossal,seo2021colossal}. The coupling between spin orientation and nodal-line degeneracy and/or chiral orbital currents have been invoked to explain the CMR in Mn$_3$Si$_2$Te$_6$ \cite{seo2021colossal,zhang2022control}. The underlying mechanisms of the observed CMR remain to be elucidated. 

The Hall resistivity $\rho_{xy}$ measured at various temperatures is displayed in Fig.~\ref{fig:2}(c) (see Supplemental Materials for more data measured at other temperatures \cite{supplemental}). Above 150 K, $\rho_{xy}(B)$ scales linearly with magnetic field with positive slops. Clearly, holelike carriers dominate at high temperatures. Nonlinear magnetic field dependence of $\rho_{xy}(B)$ develops below 150 K, suggesting multiband transport behaviours. In the temperature range between $T_\mathrm{c}$ and 150 K, $\rho_{xy}(B)$ can be well described using two holelike bands approximation \cite{chambers1952two} [see black solid lines in Fig.~\ref{fig:2}(c)]:
\begin{equation}
\begin{split}
{\rho _{xy} (B)=\frac{B}{e}\frac{(n_{h1}\mu _{h1}^{2}+n_{h2}\mu _{h2}^{2})+(n_{h1}+n_{h2})\mu _{h1}^{2}\mu _{h2}^{2}B^{2}}{(n_{h1}\mu _{h1}+n_{h2}\mu _{h2})^{2} +(n_{h1}+n_{h2})^{2} \mu _{h1}^{2}\mu _{h2}^{2}B^{2}},}
\end{split}
\end{equation}
where $n_{h1(2)}$, $\mu _{h1(2)}$ represent the carrier density, mobility of each band. Below $T_\mathrm{c}$, $\rho_{xy}(B)$ rises rapidly in small magnetic fields due to contributions from anomalous Hall effect \cite{ni2021colossal}. The simple two-band model is not sufficient to describe the Hall resistivity in the magnetically ordered state.  In Fig.~\ref{fig:2}(d), the extracted temperature-dependent carrier density and mobility of each band are shown. Above 150 K, a single hole band ($h1$) with carrier density of $n_{h1} \sim 10^{25}$ m$^{-3}$ and mobility of $\mu_{h1} \sim 10^{-3}$ m$^2$ V$^{-1}$ s$^{-1}$ dominates in electrical transport. Below 150 K, a second hole band ($h2$) becomes apparent. Compared with $h1$ , the second hole band has a lower carrier density of $n_{h2} \sim 10^{23}$ m$^{-3}$ and a higher mobility of $\mu_{h2} \sim 10^{-1}$ m$^2$ V$^{-1}$ s$^{-1}$. 

In Fig.~\ref{fig:3}, we present the Seebeck ($S_{xx}$) and Nernst ($S_{yx}$) effects of  Mn$_3$Si$_2$Te$_6$. As seen in Fig.~\ref{fig:3}(a), the Seebeck signal carries a positive sign all the way from room temperature down to 2 K. This indicates that holelike carriers play dominant roles, agreeing well with Hall resistivity. In zero magnetic field, $S_{xx}$ decreases gradually with cooling, and a kink is found at $T_\mathrm{c}$. Unlike the divergent Seebeck signal reported earlier \cite{liu2021polaronic}, $S_{xx}$ observed here approaches zero at low temperatures, in line with the third law of thermodynamics.    In 14 T, $S_{xx}$ is suppressed, which becomes apparent below about 200 K. The magneto-Seebeck effect $S_{xx}(B)$ is seen more clearly in Fig.~\ref{fig:3}(b). As one would expect, $S_{xx}(B)$ recorded between $T_\mathrm{c}$ and 150 K can also be approximated by a multiband scenario involving two holelike bands [solid lines in Fig.~\ref{fig:3}(b)] \cite{LiangCdAs,chen2022anomalous}

\begin{equation}
\begin{split}
{S_{xx}(B) =\frac{S_{h1}}{1+(\mu _{h1}B)^{2}} + \frac{S_{h2}}{1+(\mu _{h2}B)^{2}}+\frac{S_{\infty }(\mu'B)^{2}}{1+(\mu'B)^{2}},}
\end{split}
\end{equation}\\
where $S_{h1(h2)}$ and $\mu_{h1(h2)}$ are the zero-field Seebeck coefficient and carrier mobility of the corresponding hole band. $S_{\infty}$ stands for the Seebeck value when $B$ goes to infinity. The mobility obtained for each band is of the same order of magnitude as that found in the Hall resistivity. Below $T_\mathrm{c}$, $S_{xx}(B)$ shows complex behaviors especially in small magnetic fields. These features are beyond the scope of the simplified two-band approximation. The sharp cusp in $S_{xx}(B)$ near zero field is certainly linked to the metamagnetic transition as seen in magnetization measurements [see Fig.~\ref{fig:1}(c)].

At high temperatures above 200 K, the Nernst signal $S_{yx} = E_y/\nabla_x T$, as shown in Fig.~\ref{fig:3}(c), depends linearly on magnetic field. Below 150 K, nonlinear magnetic field dependence is seen in $S_{yx}(B)$, indicating the involvement of multiband transport, in agreement with the Hall effect and Seebeck effect. Notably, a step-like feature appears near zero magnetic field in the vicinity of $T_\mathrm{c}$. This feature becomes more pronounced below $T_\mathrm{c}$. The step-like behavior is very likely originated from the ANE. In conventional ferromagnets, the ANE ($S^A_{yx}$) scales linearly with magnetization ($M$), i.e., $ \left | S_{yx}^{A}  \right | =\left | N_{yx}^{A}  \right | \mu _{0}M $. Here, $N_{yx}^{A}$ is called the anomalous Nernst coefficient, which normally takes values between 0.05 and 1 $\upmu$V K$^{-1}$ T $^{-1}$ in conventional ferromagnets at 300 K \cite{guin2019anomalous,CoSnS}. As seen in Fig.~\ref{fig:3}(d), the linear scaling of Nernst signal and magnetization works nicely in Mn$_3$Si$_2$Te$_6$ below $T_\mathrm{c}$, except in low-field region. This scaling points to the dominant roles played by ANE in the Nernst signal of Mn$_3$Si$_2$Te$_6$ below $T_\mathrm{c}$. The anomalous Nernst coefficient $N_{yx}^{A}$ is found to be around  0.5 $\upmu$V K$^{-1}$ T $^{-1}$ at 70 K, which falls into the typical region spanned by conventional ferromagnets.  

\begin{figure}[t]
\includegraphics[width=\columnwidth]{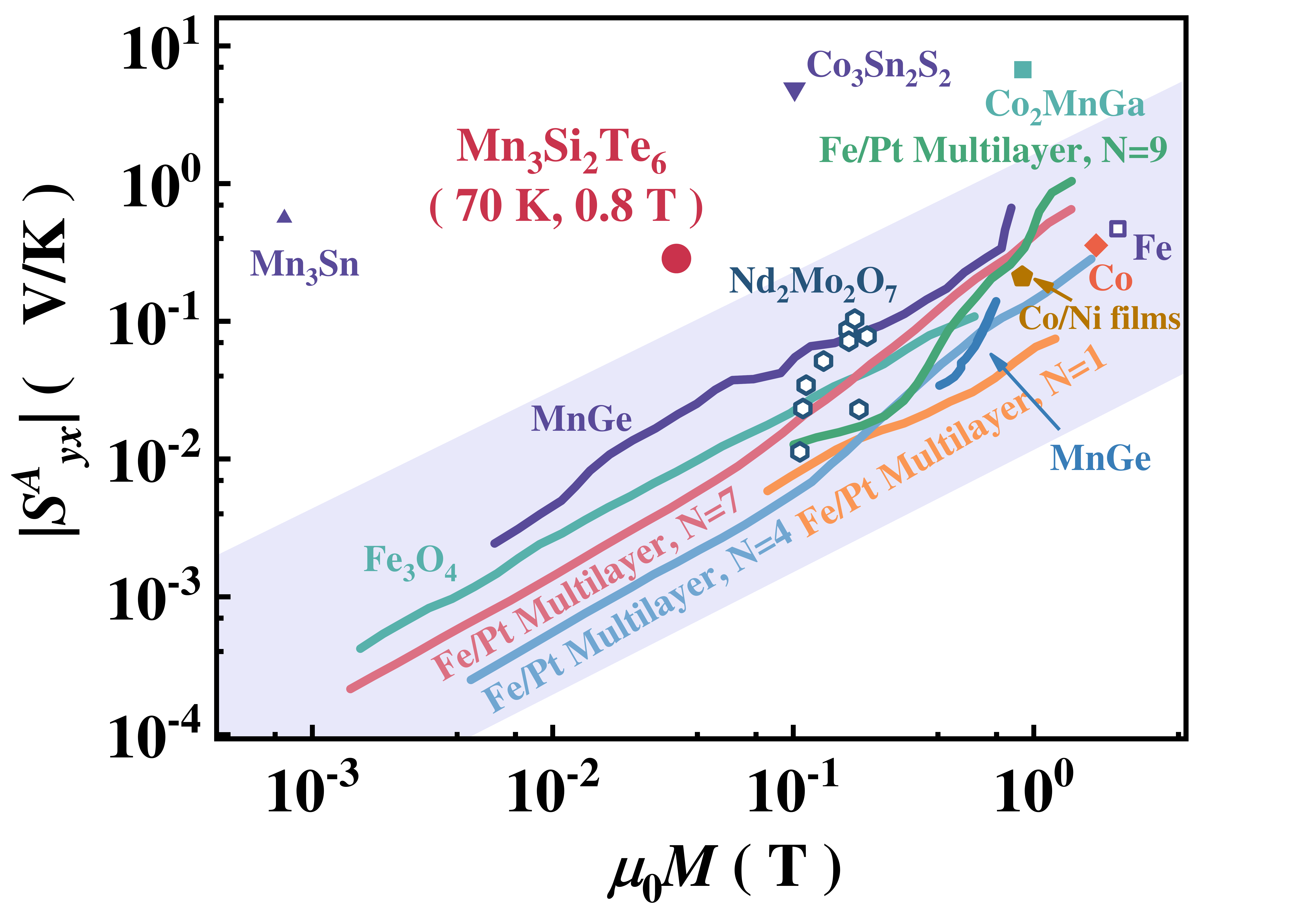}
\caption{\label{fig:4} Magnetization-dependent anomalous Nernst effect of Mn$_3$Si$_2$Te$_6$  in comparison with various conventional magnetic materials and magnetic topological materials \cite{ikhlas2017large,guin2019anomalous,ramos2014anomalous,hasegawa2015material,hanasaki2008anomalous,weischenberg2013scattering,shiomi2013topological,uchida2015enhancement}. The shade region represents the case covered by conventional magnets with $ | S_{yx}^{A}| =| N_{yx}^{A}| \mu _{0}M $, $| N_{yx}^{A}|=0.05 - 1$ $\upmu$V K$^{-1}$ T $^{-1}$. }
\end{figure}

In small magnetic fields below 3 T, the simple linear scaling between the Nernst data and magnetization fails, suggesting additional contributions. In Fig.~\ref{fig:4}, we compare the Nernst effect of Mn$_3$Si$_2$Te$_6$, conventional magnets and typical magnetic topological materials in a $ S_{yx}^{A}-M $ diagram. Clearly, the low-field ANE of Mn$_3$Si$_2$Te$_6$ locates well above the typical region covered by conventional magnetic systems. Note that the CMR effect also occurs mostly in the low-field region below 3 T [see Fig.~\ref{fig:2}(b)]. The enhanced Nernst signal in the low-field region likely has the same origin with the CMR.  In Mn$_3$Si$_2$Te$_6$, the Te $p$ orbitals can form chiral orbital states on the Te triangular lattice \cite{seo2021colossal}. Coupling of the orbital chriality and spin polarization leads to twofold nodal-line degeneracy in the Te valance bands. The nodal-line degeneracy can be lifted by finite spin-orbital coupling only when the chrial orbital angular moment is parallel to the spin. Consequently, one of the degenerated bands is raised up to $E_F$, leading to a metal-insulator transition and CMR in Mn$_3$Si$_2$Te$_6$ \cite{seo2021colossal}. Meanwhile, the topological nodal line is also pushed towards $E_F$, giving rise to sizable Berry curvature near $E_F$. As a result, the ANE gets enhanced in addition to the conventional contributions from magnetization. Berry curvature-induced large ANE has been reported in various topological magnetic materials, such as Co$_2$MnGa, Co$_3$Sn$_2$S$_2$ and Mn$_3$Sn, as also presented in Fig. \ref{fig:4}  \cite{guin2019anomalous,guin2019zero,CoSnS,DingCoSnS,ikhlas2017large,CoSnS,XuCoMnGa,Sakai2018CoMnGa}. 

Alternatively, the Nernst signal can be enhanced by ambipolar transport, in which electronlike and holelike bands are compensated \cite{Ambipolar,Ambipolar_NbSe2,Ambipolar_CsVSb}. However, the transport properties of Mn$_3$Si$_2$Te$_6$ are dominated by two holelike bands below 150 K, as we have seen in the Hall resistivity and magneto-Seebeck data. Ambipolar transport is unlikely the source of enhanced Nernst effect in small magnetic fields.  We also note that the excess contributions to Nernst signal in the low-field region are reminiscent of the topological Nernst effect,  as found in the skyrmion phases of MnGe, Gd$_2$PdSi$_3$ and Fe$_3$Sn$_2$ \cite{shiomi2013topological,hirschberger2020topological,zhang2021Topological}. In Mn$_3$Si$_2$Te$_6$, no traces of skyrmion have been found currently. Notably, chiral orbital currents (COC) running along edges of MnTe$_6$ octahedra within the $ab$ plane have been proposed in Mn$_3$Si$_2$Te$_6$ \cite{zhang2022control}. The interactions between COC-induced orbital moments, COC domains and Mn spins can also explain the unusual CMR \cite{zhang2022control}. The COC orbital moments break the time-reversal symmetry, and could serve as another prominent channel in producing large Nernst signal. This scenario has been theoretically introduced to explain the giant Nernst effect in hole-doped cuprate superconductors and heavy-fermion systems \cite{kotetes2010chirality}. Further theoretical and experimental studies are desired to verify the existence of COC, and to unravel the links between COC and Nernst effect in Mn$_3$Si$_2$Te$_6$.

In summary, we have studied electrical transport, thermoelectric Seebeck and Nernst effects in the ferrimagnetic nodal-line semiconductor Mn$_3$Si$_2$Te$_6$. The Hall effect and magneto-Seebeck effect are dominated by two holelike bands below 150 K. In the ferrimagnetic state, CMR is found when external magnetic fields are applied along the magnetic hard $c-$ axis. Moreover, ANE appears in the magnetically ordered state. In the low-field region where CMR is most pronounced,  conventional magnetization alone can not account for the observed sizable Nernst signal. The enhanced Nernst effect is most likely associated with the Berry curvature of the nodal-line topology, which may also hold the key for CMR in Mn$_3$Si$_2$Te$_6$.

\medskip

This work has been supported by National Natural Science Foundation of China (Grant Nos. 11904040, 52125103, 52071041, 12004254, 12004056, 11974065),  Chongqing Research Program of Basic Research and Frontier Technology, China (Grant No. cstc2020jcyj-msxmX0263), Chinesisch-Deutsche Mobilit\"atsprogamm of Chinesisch-Deutsche Zentrum f\"ur Wissenschaftsf\"orderung (Grant No. M-0496).

\nocite{*}

\providecommand{\noopsort}[1]{}\providecommand{\singleletter}[1]{#1}%

\end{document}